\documentclass[10pt,twocolumn,letterpaper]{article}

\usepackage{iccv}
\usepackage{times}
\usepackage{bm}
\usepackage{epsfig}
\usepackage{graphicx}
\usepackage{amsmath}
\usepackage{booktabs}
\usepackage{amssymb}
\usepackage{enumitem}
\usepackage{caption}
\usepackage{xcolor}
\usepackage{algorithm}
\usepackage{algpseudocode}
\usepackage{lipsum}
\usepackage{xspace}
\usepackage{amsthm}
\usepackage{listings}

\newtheorem{definition}{Definition}

\usepackage[pagebackref=true,breaklinks=true,letterpaper=true,colorlinks,bookmarks=false]{hyperref}

\iccvfinalcopy %

\def\iccvPaperID{1776} %

\ificcvfinal\fi

\newcommand{\patchsize}{w_T}

\newcommand{\vanilla}{MT\xspace}
\newcommand{\perm}{M2T\xspace}

\newcommand{\plotloc}[1]{\textsc{#1}}

\definecolor{code}{HTML}{785EF0}

\makeatletter
\renewcommand{\paragraph}{%
  \@startsection{paragraph}{4}%
  {\z@}{%
  1.5ex \@plus 0.1ex \@minus 0.2ex}{%
  -2ex}%
  {\normalfont\normalsize\bfseries}%
}
\makeatother

\begin{document}

\title{M2T: Masking Transformers Twice for Faster Decoding}

\author{Fabian Mentzer\\
Google Research\\
{\tt\small mentzer@google.com}
\and
Eirikur Agustsson\\
Google Research \\
{\tt\small eirikur@google.com}
\and
Michael Tschannen\\
Google Research \\
{\tt\small tschannen@google.com}
}

\maketitle
\ificcvfinal\thispagestyle{empty}\fi

\begin{abstract}
We show how bidirectional transformers trained for masked token prediction can be applied to neural image compression to achieve state-of-the-art results.
Such models were previously used for image \emph{generation} by progressivly sampling groups of masked tokens according to uncertainty-adaptive schedules.
Unlike these works, we demonstrate that predefined, deterministic schedules perform as well or better for image compression. 
This insight allows us to use masked attention during training in addition to masked inputs, and activation caching during inference, to significantly speed up our models (${\approx}4{\times}$ higher inference speed) at a small increase in bitrate. 
\end{abstract}

\section{Introduction}

Recently, transformers trained for masked token prediction have successfully been applied to neural image and video \emph{generation} \cite{chang2022maskgit, villegas2022phenaki}.
In MaskGIT~\cite{chang2022maskgit}, the authors use a VQ-GAN~\cite{esser2021taming}
to map images to vector-quantized tokens, and learn a transformer to predict the distribution of these tokens.
The key novelty of the approach was to use BERT-like~\cite{devlin2018bert} random masks during training to then predict tokens in groups during inference, sampling tokens in the same group in parallel at each inference step. Thereby, each inference step is conditioned on the tokens generated in previous steps.
A big advantage of BERT-like training with grouped inference versus prior state-of-the-art is that considerably fewer steps are required to produce realistic images (typically 10-20, rather than one per token).

These models are optimized to minimize the cross entropy between the token distribution $p$ modeled by the transformer and 
the true (unknown) token distribution $q$, as measured via negative log likelihood (NLL).
As is known from information theory, this is equivalent to the bit cost required 
to (losslessly) store a sample drawn from $q$ with a model $p$~\cite{Yang2022a}.
Indeed, any model $p$ that  predicts an explicit joint distribution over tokens in a deterministic way can be turned into a \emph{compression} model by using $p$ to entropy code the tokens, rather than sampling them. 

\begin{figure}[t]
    \centering
    \includegraphics[width=0.9\linewidth]{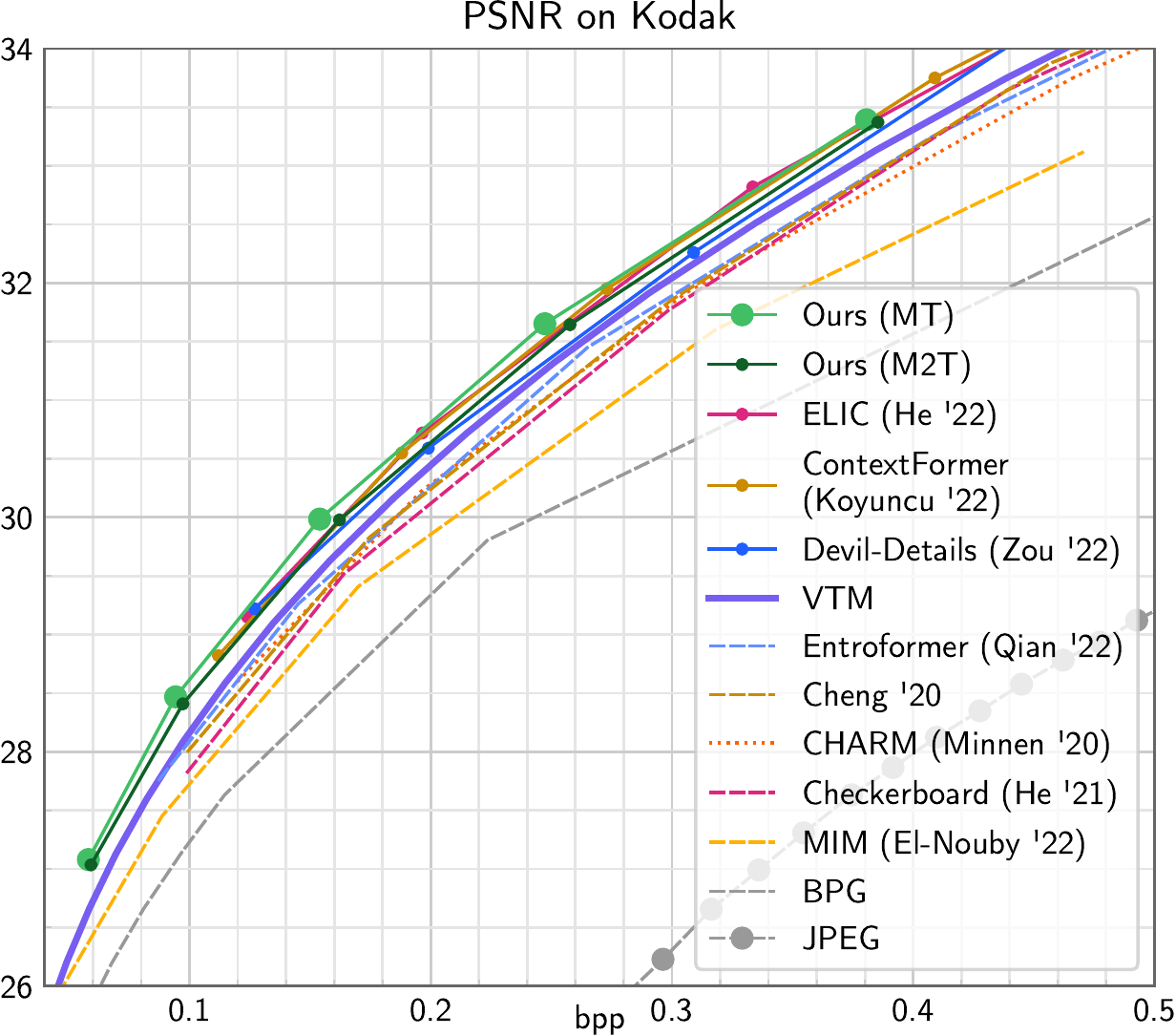}\vspace{-1.3ex}
    \caption{\label{fig:rd}Rate distortion results on Kodak. 
    Our \emph{\vanilla} outperforms the prior state-of-the-art ELIC \cite{he2022elic}; 
    \emph{\perm} only incurs a small reduction in rate-distortion performance compared to MT while running about $4\times$ faster on hardware (see Fig.~\ref{fig:speed})\vspace{-1.9em}}
\end{figure}

\begin{figure*}[ht]
    \centering
    \includegraphics[width=\linewidth]{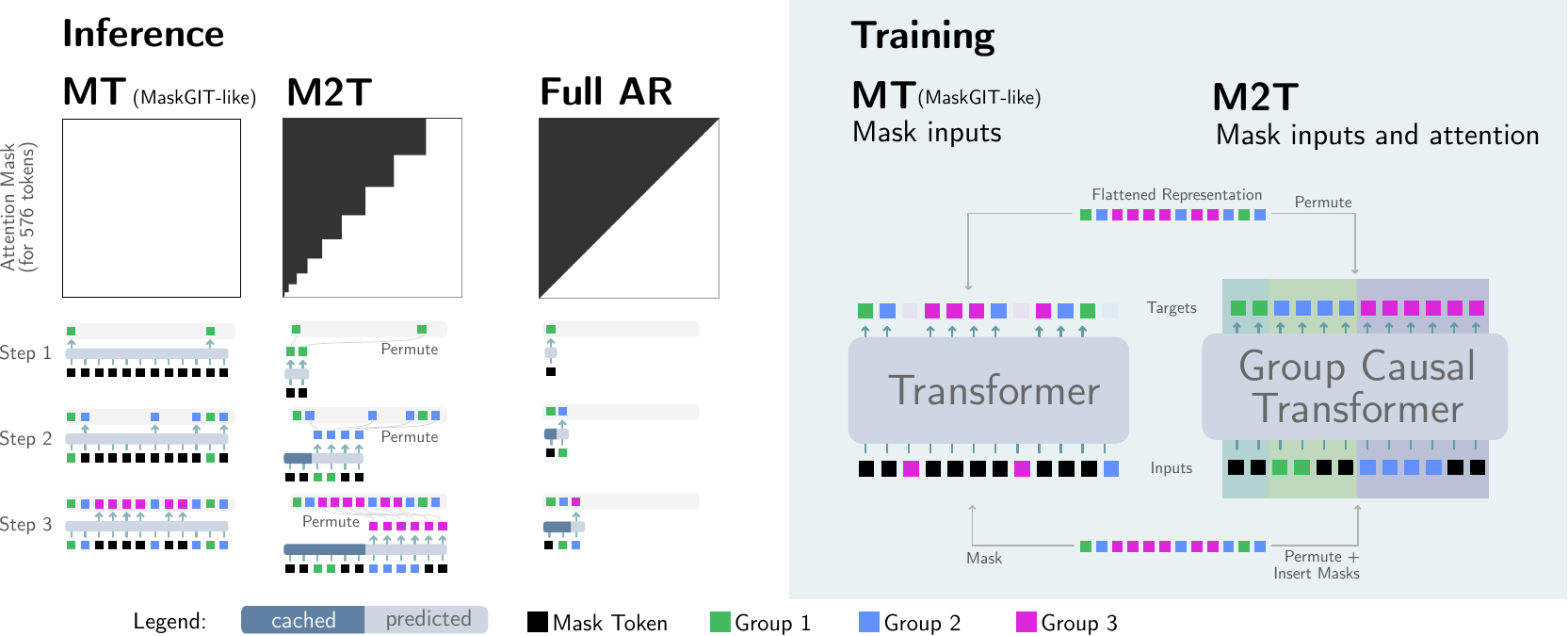}
    \caption{
    \label{fig:method} 
    \plotloc{Left}: 
    At the top we show the attention masks, at the bottom the first three inference steps. 
    We first show the MaskGIT-like approach \textbf{\vanilla}, where the attention is not masked, and the same number of tokens is fed in each step.
    In our \textbf{\perm} approach, both the attention and the input is masked, and we uncover the input one group at a time.
    The causal masks used in the attention allow us to cache activations, \ie, \textit{for shaded blue regions we can cache}. Together with the fewer tokens fed in each (but last) step, this significantly speeds up the model.
    Finally, we show the fully autoregressive (\textbf{Full AR}) approach for reference, as employed, \eg, in the standard transformer decoder~\cite{vaswani2017attention}. Our \perm approach is a generalization that allows for group sizes greater than one.
    \plotloc{Right}: Training, shown for 12 tokens only. 
    Note that each input group corresponds to the previous output group with additional mask tokens to align groups of different sizes.
    The group causal transformer is using attention masking
    (see Sec.~\ref{sec:masking:perm} for details). \vspace{-1em}
    }
\end{figure*}

Motivated by this, we aim to employ masked transformers for neural image compression. 
Previous work has used masked and unmasked transformers in the entropy model for video compression~\cite{xiangmimt,mentzer2022vct} and image compression~\cite{qian2022entroformer,koyuncu2022contextformer,el2022image}.
However, these models are often either prohibitively slow~\cite{koyuncu2022contextformer},
or lag in rate-distortion performance~\cite{qian2022entroformer,el2022image}.
In this paper, we show a conceptually simple
transformer-based approach that is state-of-the-art in neural image compression,
at practical runtimes. The model is using off-the-shelf transformers, and does not rely on special positional encodings or multi-scale factorizations, in contrast to previous work.
Additionally, we propose a new variant 
combining ideas from MaskGIT-like input-masked transformers and fully autoregressive attention-masked transformers.
The resulting model masks both the input and attention layers, and allows us to substantially improve runtimes at a small cost in rate-distortion.

To train masked transformers, the tokens to be masked in each training step are usually selected uniformly at random. 
During inference, the models are first applied to mask tokens only,
predicting a distribution for every single token.
A sample is then drawn from this distribution and a subset of tokens is uncovered at the input (see Inference/\vanilla in Fig.~\ref{fig:method}). This step is repeated until no mask tokens remain.
Two important questions arise:
i) How many tokens do we sample in every step, and ii) which spatial locations do we chose.
In MaskGIT, an instance-adaptive scheme is used for (ii), \ie, every sampled image will have a different schedule of locations.
In this work, we show that in terms of NLL (and thus bitrate), a fixed schedule performs just as well.

This allows us to generalize ideas used in fully autoregressive transformer decoders like the original model proposed by Vaswani~\etal~\cite{vaswani2017attention},
bridging between fully autoregressive and MaskGIT-like transformers, as follows:
In autoregressive models, the input sequence is shifted by one token to the right, causing the outputs to align in a casual way, \ie, the $i$-th output is trained to predict the ($i-1$)-th input (see ``Full AR'' in Fig.~\ref{fig:method}).
This can be thought of as a ``group-autoregressive'' schedule with group size equal to 1.
We generalize this idea to group sizes ${>}1$:
As shown in Fig.~\ref{fig:method} (``\perm''), we permute the input such that we can uncover it group by group from left to right,
and permute the targets such that each group at the input predicts the next group at the output.
To accommodate a sequence of increasing group size (which leads to the best generation/compression performance in practice) we insert mask tokens at the input to pad the $i-1$-th group to the length of the $i$th group.
During inference, this allows us to run the transformer first on very few tokens, and then more and more.
In contrast, MaskGIT-like transformers always feed the same number of tokens (some are masked), corresponding to the full image. %
We apply this idea to neural image compression, and show that our approach reduces the average compute per output token as it processes fewer tokens in total, at a small cost in bitrate.

Our core contributions in this paper are \emph{two models}:

\paragraph{Model 1 (\vanilla)} 
A vanilla MaskGIT-like transformer that obtains state-of-the art neural image compression results. In contrast to previous work, we use a conceptually clean approach relying on standard transformers applied to tiles;
our method does not require a multi-scale model (``hyperprior''), and we can span a large bitrate regime by using scalar quantization.

\paragraph{Model 2 (\perm)} We show how \vanilla can be sped up by masking the transformer twice: both at the input and in the attention layers.
As visualized in Fig.~\ref{fig:method}, the model is faster because it is applied to fewer tokens and because the attention masks make the transformer causal, allowing for caching.
Together, this leads to to $2.7{\times}-4.8{\times}$ runtime improvements as measured on accelerators, vs.\ a MaskGIT-like model.

\section{Related Work} \label{sec:relwork}
Lossy neural image compression is an active field of research, with advancements being made on two fronts: 
Entropy models (how to losslessy code a lossy, quantized representation of the image) and 
transforms (how to encode/decode the representation from/to pixels). 

On the transform side, earlier methods used residual blocks~\cite{theis2017lossy} and Generalized Divisive Normalization (GDN)~\cite{balle2016end}, but more recently residual blocks with simplified attention~\cite{cheng2020learned, he2022elic} and window-based transformers~\cite{zhu2021transformer, zou2022devil} have been employed in state-of-the-art methods. Another line of work tackles generative compression where the synthesis transform is trained to generate texture and low-level detail at low rates \cite{tschannen2018deep, blau2019rethinking, agustsson2019extreme, mentzer2020high, el2022image}.

On the entropy modeling side, most methods have built on top of the hyperprior~\cite{balle2018variational} paradigm, where the representation is modelled with a (two-scale) hieararchical latent variable model.
Further improvements include channel autoregression, ``CHARM''~\cite{minnen2020channel}, and checkerboard modeling~\cite{he2021checkerboard}, employing a limited number of auto-regression steps over space and/or channels.
Fully autoregressive models are sometimes used in the literature~\cite{mentzer2018conditional,minnen2018joint, cheng2020learned,koyuncu2022contextformer} to further reduce bitrates, but their prohibitively slow runtimes render them less practical (often requiring minutes to decode high-resolution images). 

Recently, transformers have been investigated both for the entropy models and the transforms. 
Qian~\etal~\cite{qian2022entroformer} 
fuse together an autoregressive transformer and a hyperprior~\cite{balle2018variational} (using transformer encoders instead of CNNs there as well).
They introduce a top-k scheme in the attention layer and a special relative positional encoding to handle arbitrary resolutions.
El-Nouby~\etal~\cite{el2022image} use a Masked Image Model (MIM) combined with a Product Quantization (PQ) variant of VQ-VAE. While the approach is promising for extreme compression, in terms of rate-distortion the method is lagging behind state-of-the-art significantly. 
Konyuncu~\etal~\cite{koyuncu2022contextformer} propose a transformer based entropy model that is fully auto-regressive over the spatial and channel dimensions, which leads to prohibitively slow decode times (10+ minutes for a 4K image).
Other works~\cite{zhu2021transformer, zou2022devil} have explored the use of window-based transformers for the synthesis transform.

For neural video compression, VCT~\cite{mentzer2022vct} demonstrated strong results with a temporal transformer for entropy modeling and more recently~\cite{xiangmimt} combined masked image transformers with multi-scale motion context (via optical flow + warping) to obtain state-of-the-art results.

\section{Method}

\subsection{Overview}

\begin{figure}
    \centering
    \includegraphics[width=\linewidth]{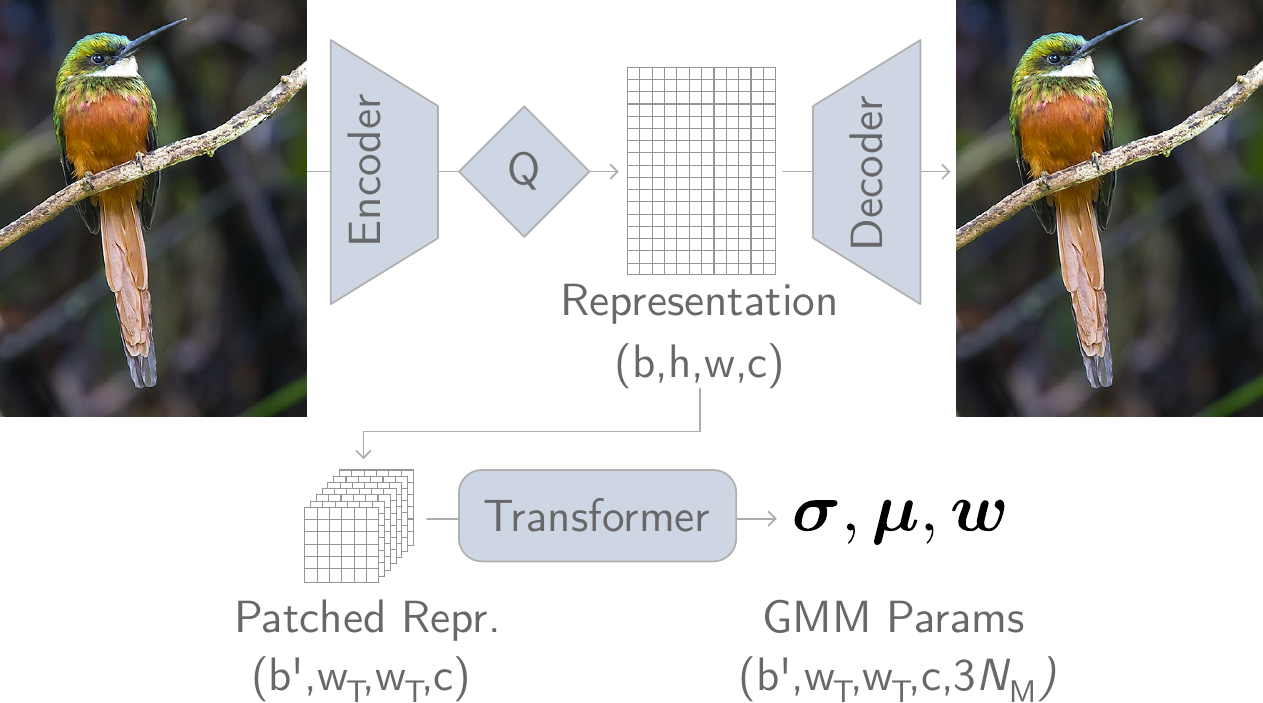}
    \caption{\label{fig:arch} Architecture overview. The Encoder maps a batch of input images to a discrete representation of shape $(b,h,w,c)$. This representation is then split into patches of size $w_T$ (folded into the batch dimension, so that $b' = b \cdot hw/\patchsize^2$). These are each entropy-coded independently (and possibly in parallel) using the distributions predicted by \vanilla or \perm, which is parameterized by a GMM with $N_\text{M}{=}3$ mixtures.
    }
\end{figure}

A high level overview of our approach is shown in Fig.~\ref{fig:arch}.
Given an $H{\times}W$ image, we apply an encoder $E$ to obtain a features of shape $(\lceil H/16 \rceil, \lceil W/16 \rceil,c)$, which we quantize element wise (scalar quantization), following many previous works~\cite[\dots]{balle2016code,mentzer2018conditional,cheng2020learned,he2022elic,minnen2018joint}, 
yielding the representation $y = Q(E(x))$.
From $y$, we can get a reconstruction with a decoder $D$, $\hat x = D(y)$.

Since in general, $\hat x \neq x$, we call this a \emph{lossy} auto-encoder. We can turn it into a lossy image compression scheme by storing $y$ to disk losslessly.
For example,  we could use the naive way of storing every element in $y$ independently with an \texttt{int32}, resulting in a method that uses $32c/16^2$ bits per pixel (bpp). This results in a very poor compression ratio, so instead,
we follow previous work in predicting a (discrete) distribution $P(y)$,
to then use entropy coding to store $y$ to disk using $\approx \sum_i -\log_2 P(y_i)$ bits (intuitively, more likely symbols should be stored with fewer bits).
We refer to previous work on the theoretical background, see, \eg, Yang and Mandt~\cite{Yang2022a} and Balle~\etal~\cite{balle2020nonlinear}.
Here, we shall use a 
masked transformer to model $P$.

\subsection{Autoencoder and Tokenization}

\begin{figure*}[ht]
\centering
\includegraphics[width=\textwidth]{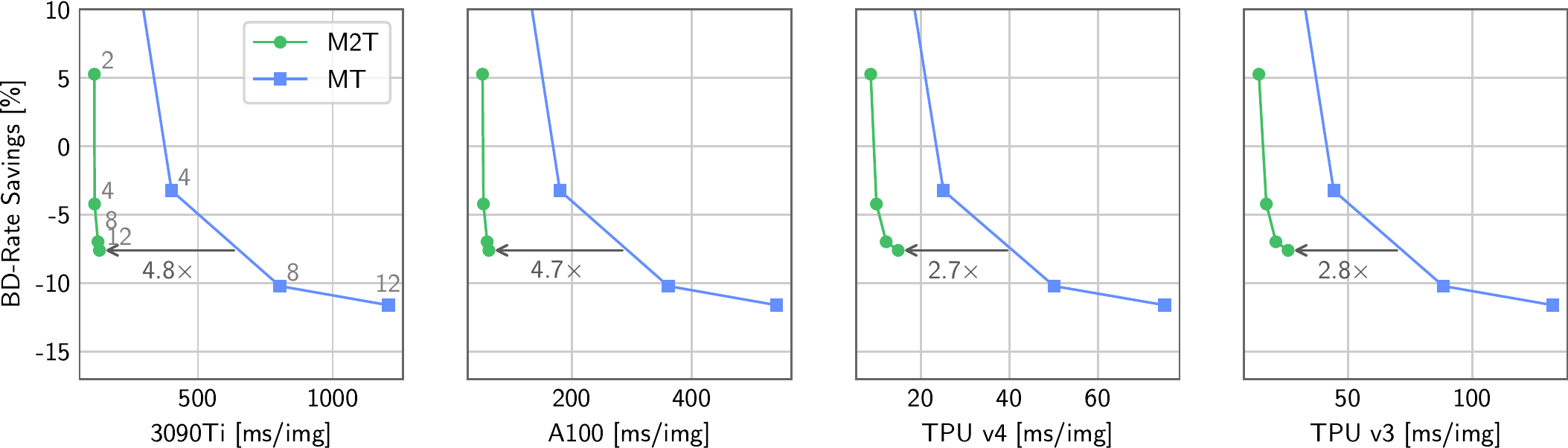}
\caption{\label{fig:speed}
We compare our models \vanilla and \perm in terms of speed vs.\ rate savings over VTM (lower is better).
We compare milliseconds per image on various platforms,
where ``image'' is a large $1500{\times}2000$ pixel image.
TPU numbers are obtained by using 4 chips in parallel.
The trade-off between rate savings and speed is controlled by adjusting the number of inference steps $S$, which we annotate for the first plot.
We see that at the same rate savings, \perm is $2.7{\times}-4.8{\times}$ faster.
Both models start to saturate in terms of rate savings at $S=8$.
}
\end{figure*}

Our main contributions lie in how we model $P$, so for the autoencoder we use the convolutional ELIC encoder/decoder proposed by He~\etal~\cite{he2022elic},
with 256 channels for all layers except for the last layer of the encoder which predicts the $c$-dimensional representation. We use $c{=}192$ throughout.
$E$ downscales by a factor $16$, and we use $h{=}\lceil H/16\rceil, w{=}\lceil W/16\rceil$ as shorthand.\footnote{%
If the dimensions of an image do not divide by 16 during inference, we pad the image, calculate the padded reconstruction and bitrate, and unpad at the output, following previous work~\cite[\dots]{minnen2018joint,cheng2020learned,mentzer2020high}
}
To get gradients through the quantization operation, 
we rely on 
straight-through estimation (STE)~\cite{theis2017lossy,minnen2020channel}.

Similar to previous work applying transformers to compression~\cite{xiangmimt,mentzer2022vct,koyuncu2022contextformer},
we do \emph{not} consider each of the $h\cdot w\cdot c$ elements in this representation as a token, since this would yield infeasibly long sequences for transformers (\eg, a $2000{\times}2000$px image turns into a representation with $125{\times}125{\times}192=3M$ symbols).
Instead, we group each $1{\times}1{\times}c$ column into a ``token'', \ie, we get $hw$ tokens of dimension $c$ each. 

\subsection{Transformer}
We use a standard transformer encoder in the pre-norm setup (see, \eg,~\cite[Fig.~1]{dosovitskiy2020image} and~\cite{xiong2020layer}) with the Base (``B'') config~\cite{devlin2018bert,dosovitskiy2020image} (12 attention layers, width 768, and MLPs with hidden dimension 3078).
We apply two compression-specific changes:
since our input is a vector of $c$ scalar-quantized integers, we cannot use the standard dictionary lookup-based embedding (as the vocabulary size is theoretically infinite). 
Instead, we normalize the vectors by dividing with a small constant $\delta{=}5$ and apply a dense layer shared across tokens to function as the ``embedding layer''.
Similarly, at the output, we cannot simply predict a finite number of logits. Rather, we follow the standard practice in neural compression and pixel-autoregressive generative modeling to model each entry of a token using a continuous, parametrized distribution, which is then quantized to a PMF as described below. Inspired by~\cite{salimans2017pixelcnnpp,cheng2020learned}, we use a mixture of Gaussians (GMM) with $N_\text{M}=3$ mixtures, each parameterized by a mean $\mu$, scale $\sigma$, and weight $w$.

\paragraph{Patched inference} For standard transformers, a positional embedding is typically learned for every input token, and we also apply this.
This means that these models are not applicable to arbitrary resolutions during inference without carefully adapting the positional embedding, which often involves finetuning on the target resolution.
However, for image compression, datasets of widely varying image size are the norm.
To reconcile this, we use a simple solution: 
we apply the transformer on patches of $\patchsize{\times}\patchsize$ tokens. 
We use $\patchsize=24$ since this corresponds to full representation size during training (we use 384px crops during training, yielding $h=w=24$).
Since we use the transformer for \emph{losslessly} coding the representation, we do not see any boundary artifacts from this technique.
The only downside is that some correlations across patches are not leveraged to drive down the bitrate even further.
Concretely, this implies the following flow of tensors shown in Fig.~\ref{fig:arch} during inference.

We emphasize the simplicity of our proposed scheme, using off-the-shelf transformers in a patched manner.
In contrast to, \eg, Entroformer~\cite{qian2022entroformer}, we do not have to adapt the attention mechanism or use a relative positional embedding. This means that our approach will benefit from future research into speeding up standard transformers.

\begin{figure*}
    \centering
    \begin{tabular}{rl}
    \hfill\begin{minipage}[c]{0.09\linewidth}%
    Entropy\\
    \hfill$\alpha=2.2$
    \end{minipage}&
    \begin{minipage}[c]{0.7\linewidth}%
    \includegraphics[width=\linewidth]{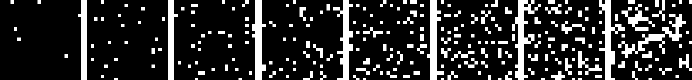}\vspace{.8ex}
    \end{minipage}\\
    \hfill\begin{minipage}[c]{0.09\linewidth}%
    Random\\
    \hfill$\alpha=2.2$
    \end{minipage}&
    \begin{minipage}[c]{0.7\linewidth}%
    \includegraphics[width=\linewidth]{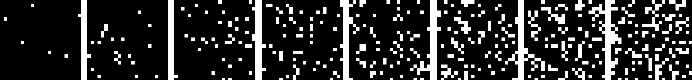}\vspace{0.8ex}
    \end{minipage}\\
    \hfill\begin{minipage}[c]{0.09\linewidth}%
    QLDS\\
    \hfill$\alpha=2.2$
    \end{minipage}&
    \begin{minipage}[c]{0.7\linewidth}%
    \includegraphics[width=\linewidth]{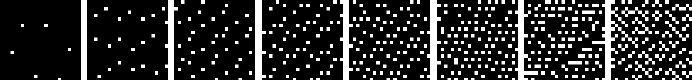}
     \end{minipage}\\
    \end{tabular}
    \vspace{2ex} \\
    \hspace{3.3em}\includegraphics[width=0.74\linewidth]{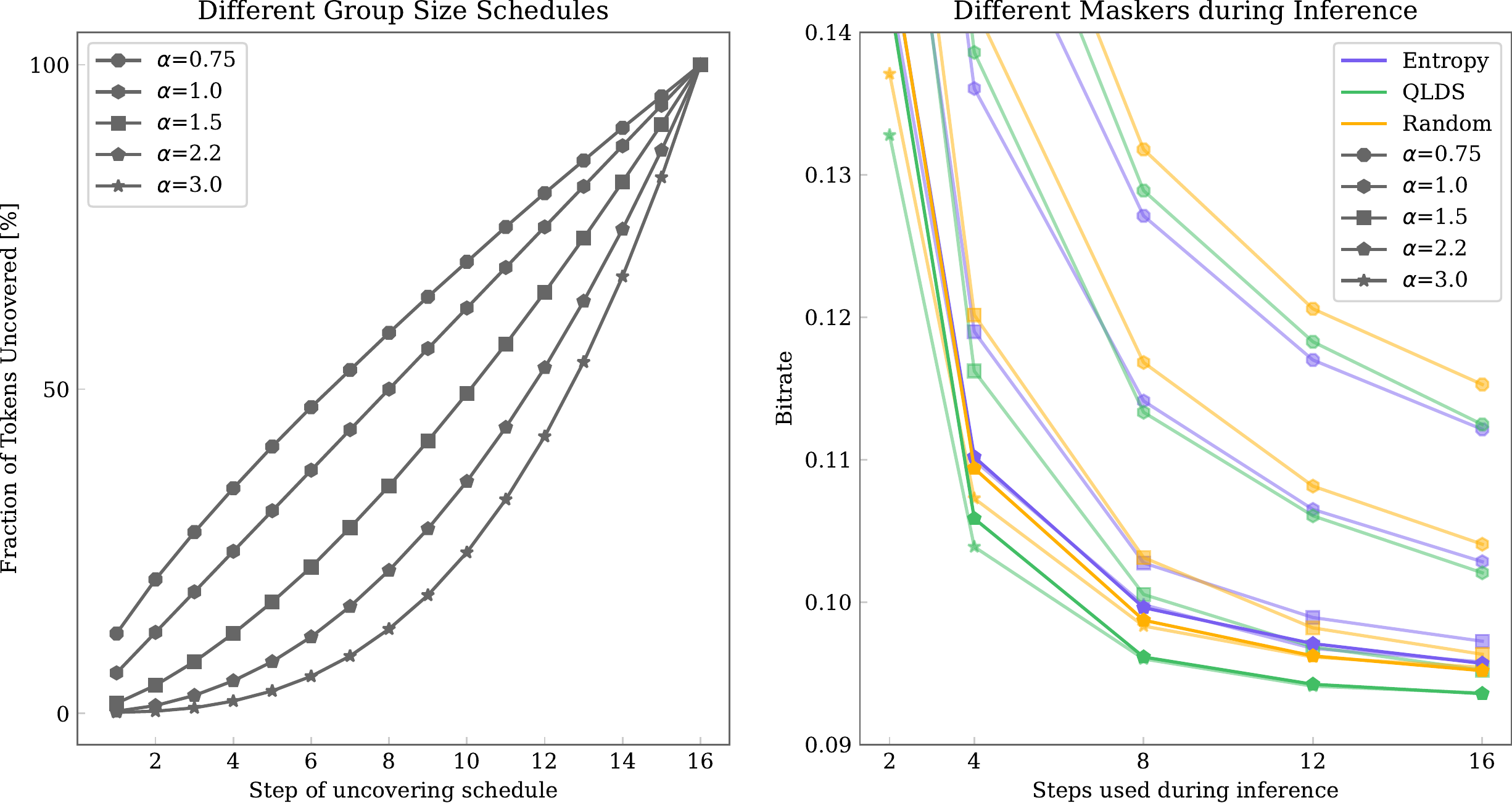}
    \caption{\label{fig:schedules}%
    \plotloc{Top}: Different location schedules shown for 8 steps and $\alpha=2.2$. Note that \emph{Entropy} is instance adaptive and we show it for one instance.
    \plotloc{Bottom Left}: Visualizing how different $\alpha$ look in terms of how much is uncovered after each step (cumulative).
    \plotloc{Bottom Right}: Resulting inference bitrate of the various $\alpha$, shown for different location schedules and different inference steps. Each point is \textbf{the same \vanilla model} evaluated with a different schedule as parameterized by the triplet ($\alpha$, inference steps, location schedule), where $\alpha$ is shown via the marker.}
\end{figure*}

\subsection{Masking Schedules}\label{sec:masks}

We consider various masking schedules in this work.
A masking schedule is a sequence of masks $\mathcal M = \{\bm M_1, \dots, \bm M_S\}$, where $S$ is the number of masks (or, equivalently, inference steps), and each tensor $\bm M_i$ is a binary mask tensor of length $\patchsize^2$. $\bm M_i[j]=1$ indicates that the $j$-th token is predicted and \emph{uncovered} at step $i$.
As outlined in the introduction, there are two important axes when building $\mathcal M$ besides the number of masks $S$:
\begin{enumerate}[leftmargin=*]
    \item \textit{Group Size Schedule}: How many token are uncovered in each step, \ie, what is $\sum_j \bm M_i[j]\,\forall i$.
    \item \textit{Location Schedule}: Which tokens are chosen to be uncovered, \ie, which indices in each $\bm M_i$ are set to 1.
\end{enumerate}

We parameterize the ``group size schedule'' via the cumulative number of tokens that are uncovered after $x$ steps using a strictly monotonically increasing function $f(x)$.
Motivated by MaskGIT, we limit ourselves to a power schedule, \ie, $f(x) = N_{S,\alpha} x^\alpha$, where $\alpha$ controls how fast we uncover, and $N_{S,\alpha}$ normalizes such that we uncover all $\patchsize^2$ tokens in $S$ steps.
Fig.~\ref{fig:schedules} shows $f(x)$ for some $\alpha$.

For ``location schedules'' we consider three different options,
visualized at the top of Fig.~\ref{fig:schedules}.
Again motivated by MaskGIT, we start with an \textbf{entropy}-based schedule. 
MaskGIT uses a schedule where in the $i$-th step, the model is applied to the current input, a distribution $p_j$ is predicted for every masked token, and a value $x_j$ is sampled for every masked location $j$. A ``confidence score'' of $x_j$ is obtained as $p_j(x_j)$ and a number of tokens (governed by the group size schedule) with the highest confidence score is retained. This also determines the masked locations of the next step $i+1$.
For compression, since we aim to produce short bitsteams, and the bitrate is a function of the predicted entropy,
we follow~\cite{xiangmimt} and adapt this schedule to our use case by retaining tokens with the lowest entropy instead of the confidence score.

The second schedule is called \textbf{random}, where we fix a seed and sample locations at uniformly at random (with a fixed seed), motivated by the fact that this mimics the training distribution of mask locations. %

Our last schedule is a novel schedule proposed in this paper, \textbf{QLDS} (``quantized low-discrepancy sequence''), which is loosely motivated by information theory:
We note that at every step $i$, we entropy code the tokens in the $i$-th group in parallel, and conditionally on the tokens of all previous groups (this is possible, as these tokens will be available in the $i$-th decoding step).
Hence, to get good prediction of all available at tokens in the $i$-th group, the mutual information between the $i$-th group and all previous groups should be maximized.
At the same time, all tokens \emph{within} a group are encoded in parallel, and we can thus not leverage their mutual information, meaning the schedule should minimize the mutual information within groups.
For images we can use distance in pixel space as a proxy for mutual information, since we expect nearby pixels to be more correlated than pixels far apart.
Intuitively, this implies that tokens within a given group should be far from each other spatially, and at the same time close to tokens in previous groups.

\begin{figure}
    \centering
    \includegraphics[width=0.7\linewidth]{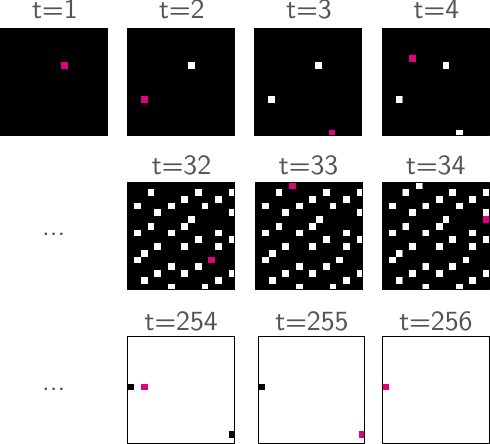}
    \caption{\label{fig:qlds} Visualizing how the quantized low-discrepancy sequence (QLDS) fills a $16{\times}16$ window in a regular fashion. We split this into a sequence of $S$ masks for our model (see Fig.~\ref{fig:schedules} for $S=8$).}
\end{figure}

To this end, we use low-discrepancy sequences (LDS)~\cite[Ch.~2]{kuipers2012uniform}.
These are pseudo-random sequences that minimize the ``discrepancy'' for any subsequence, meaning among other things that when the sequence is cut off at an arbitrary index $i$, all elements up to $i$ are close to evenly distributed (see Sec.~\ref{app:sec:lds} for a formal definition).
An LDS in 2D is given by a sequence of points $X={\bm x_1, \dots,\bm x_N}$.
This can be turned into a masking schedule by specifying $K$ group sizes that sum to $N$, and then simply splitting $X$ into $K$ groups.
The fact that $X$ is an LDS implies the desired properties mentioned above, \ie, 
all points in a group are far from each other, while at the same time merging all groups up to a certain step yields a set of points that near-uniformly cover the space.
We use an LDS proposed by Roberts~\cite{roberts2018unreasonable}, described in Sec.~\ref{app:sec:lds}, visualized in Fig.~\ref{fig:qlds}.

\subsection{Masking Model 1: \vanilla}

For our MaskGIT-like model, \vanilla, 
we use masked transformers similar to what was proposed in previous language and image generation work~\cite{devlin2018bert,chang2022maskgit,chang2023muse}.
\paragraph{Training}
Given the representation $y=E(x)$, we randomly sample a mask $\bm M$ for every batch entry, which is a binary vector of length $\patchsize^2$, where 5-99\% of the entries are \texttt{1}. The corresponding entries in $y$ are masked, which means we replace them with a special mask token (this is a learned $c$-dimensional vector).
The resulting tensor, $y_{\bm M}$, is fed to the transformer, 
which predicts distributions of the tokens. Each distribution is factorized over the $c$ channels.
We only consider the distributions corresponding to the masked tokens to compute the loss, \ie,
\begin{equation}
    \mathcal{L}_\text{\vanilla}=\mathbb{E}_{y\sim p_y,u\sim \mathcal U_{\pm0.5}} \big[%
    \sum_{%
    \substack{
    i \in \{1,..,\patchsize^2\},\\\bm M[i]=1
    }} 
      \hspace{-2ex} -\log_2 p(y_i + u | y_{\bm M})
        \big],\label{eq:loss:vanilla}
\end{equation}
where we use additive i.i.d.\ noise to simulate quantization during training~\cite{Yang2022a}.
Here, 
we use the standard trick (\eg~\cite{Yang2022a}) of integrating the continuous distribution $p$ produced by the model on unit-length intervals, 
to obtain $P(y)=\int_{y-1/2}^{y+1/2}p(u)du,y\in\mathbb Z$.

\paragraph{Inference} 
For inference, we apply the model $S$ times following one of the schedules outlined in Sec.~\ref{sec:masks}.
In the first iteration, we only feed mask tokens, then we entropy code the tokens corresponding to $\bm M_1$, uncover them at the input, and repeat until all tokens have been entropy coded. This is detailed in 
Fig.~\ref{fig:method} (left) and Alg.~\ref{alg:send:vanilla} in the Appendix.
In Fig.~\ref{fig:sampling} we qualitatively visualize how the prediction gets more confident in each step as more tokens are uncovered. 

\subsection{Masking Model 2: \perm} \label{sec:masking:perm}

As we shall see in Sec.~\ref{sec:res:ablations},
we can use a deterministic schedule for inference without hurting bitrate in \vanilla.
This motivates our fast model that masks \textbf{twice}: once at the input, once in the attention, called \textbf{\perm} (see Fig.~\ref{fig:method}). 

Recall that fully autoregressive transformer decoders like the original approach by Vaswani~\etal~\cite{vaswani2017attention} use a diagonal attention mask during training to enforce causality.
We generalize this idea here.
Given a sequence of masks $\mathcal M$, we construct i) a permutation of the input, ii) attention masks, iii) a permutation of the targets, which together allow us to get the complete token distribution with a single forward pass during training, and, \emph{crucially} allow us to do fast inference.

As visualized in Fig.~\ref{fig:method}, 
we can (i) form the permuted inputs by constructing $|\mathcal{M}|$ groups, 
where the $i$-th group consists of the tokens in group $\bm M_{i-1}$ followed by mask tokens to pad the subsequence to length $\sum \bm M_i$.
$\mathcal M$ also induces (ii) an attention mask $\bm A$, a ``block triangular'' matrix (see Fig~\ref{fig:method}) which ensures causal dependence structure across groups.
Finally, (iii) the permutation of the targets is simply putting tokens of the same group next to each other.

\begin{figure*}[ht]
\centering
\includegraphics[width=\linewidth]{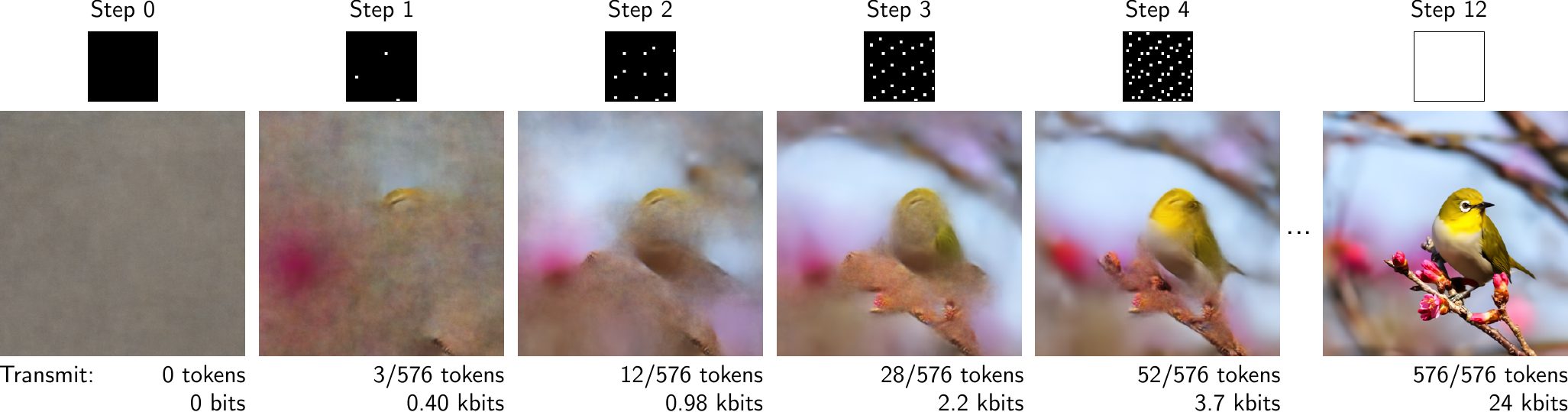}
\caption{\label{fig:sampling}Visualizing the uncertainty of the model
for a QLDS masking schedule with $\alpha{=}2.2$ and $S{=}12$. 
Above each image we show the \emph{cumulative} mask at the corresponding step, where transmitted token locations are indicated with a white dot. We show the cost for storing these tokens to disk in kilobits (kbits). 
To visualize the uncertainty of the model, we sample the remaining (non-transmitted) tokens 50 times and average the corresponding decoded images. 
We see how the QLDS schedule leads to a coarse-to-fine transmission of the data.
}
\end{figure*}

We emphasize that mask tokens at the input enable nonlinear schedules where the current step predicts more tokens than the previous step, by simply padding the previously predicted/decoded tokens at the input to the length of the output of the current step. %
Further, masking the attention turns the model into a causal transformer which allows teacher forcing during training \cite{williams1989learning}, \ie, all steps can be trained simultaneously. This also enables caching at inference time.

We highlight that this scheme is a generalization of full autoregressive training with attention masks: 
We recover it with a sequence of masks $\mathcal M = \{ [1, 0, \dots], [0, 1, \dots], \dots , [\dots, 0, 1]\}$ that uncover the latent in raster scan order. 
Using this, we obtain groups of size 1, and the standard triangular $\bm A$ (see Fig.~\ref{fig:method}, ``Full AR'').
According to our algorithm outlined above, we only insert a single mask token at the start of the input. This corresponds to the \texttt{START} token typically used with fully autoregressive models.
We do not study this setting here, since running fully autoregressive transformers for compression leads to impractically long decoding times.

\paragraph{Training} 
During training, we apply the components above: (i) we permute the input,
obtaining $y_\text{in}$, (ii) feed it to the transformer masked with attention masked by $\bm A$,
and (iii) get the permuted output $y_\text{out}$, yielding 
\begin{equation}
    \mathcal{L}_\text{\perm} = \mathbb{E}_y \big[%
        -\log_2 p(y_\text{out} | y_\text{in})
        \label{eq:loss:perm}
    \big].
\end{equation}
In contrast to the loss for \vanilla (Eq.~\ref{eq:loss:vanilla}),
this loss corresponds to the bitrate required to compress the full $y$.

\paragraph{Inference}
For inference, we feed slices of the input as shown on the left of Fig.~\ref{fig:method}. We cache activations for the tokens we previously fed, which works thanks to the causality we induce during training with $\bm A$.
Note that default attention caching implementations usually is only valid for the fully autoregressive case, and we thus implemented our own Flax~\cite{flax2020github} attention caching. Code for this is shown in App.~\ref{app:sec:flax}. 
We further note that \vanilla uses the flax \texttt{MultiHeadDotProductAttention} without modification as it does not invove attention masking.

\subsection{Loss}

We train the autoencoder and transformer jointly end-to-end,
minimizing the rate-distortion trade-off
$r(y) + \lambda d(x, \hat x)$.
We use either 
$\mathcal{L}_\text{\vanilla}$ or
$\mathcal{L}_\text{\perm}$ for $r(y)$ and MSE for $d$.
The hyperparameter $\lambda$ controls the trade-off between the bitrate and distortion.%

\section{Experiments}

\paragraph{Models}
We call our base transformer model without attention masking \textbf{\vanilla},
and our model that masks twice \textbf{\perm}.
They share all hyperparameters. We explore $S{=}\{2, 4, 8, 12\}$.
For the main results, we fix $\alpha{=}2.2$, and use $S{=}12$.
In terms of rate distortion, we compare to various models listed in Sec.~\ref{sec:relwork}. We run VTM 17.1~\cite{vtm17}, the state-of-the-art non-neural codec, with the same settings as previous work
~\cite[Sec. A.2]{agustsson2022multi}.%

\paragraph{Training} 
We train our models from scratch end-to-end, including the autoencoder $E,D$.
Our training data consists of a set of 2M high-resolution images collected from the Internet,
from which we randomly sample $384{\times}384$ crops with batch size 32.
We optimize the training loss for five values of $\lambda \in {2^i : i \in \{-4, ..., 0\}}$, training for 1M steps for each $\lambda$.
We use ``$\lambda$ warmup'' where we set $\lambda$ $10{\times}$ higher for the first 15\% of training.
We set the base learning rate to $10^{-4}$, and use linear warmup for 10k steps, keep the learning rate constant until 90\% of the training is completed, and then drop it by $10{\times}$. This tends to boost PSNR and is commonly done~\cite{minnen2020channel,mentzer2022vct}.
Sec.~\ref{app:sec:details} shows implementation details.

\paragraph{Test Data}
We use the common Kodak~\cite{kodakurl} dataset to evaluate our model.  This is a dataset of 24 images, each of resolution $512{\times}768$ or $768{\times}512$. We also evaluate on the full CLIC2020 dataset, which consists of 428 images of up to $2000{\times}1000$px.\footnote{\url{www.tensorflow.org/datasets/catalog/clic}}
We report bits-per-pixel (bpp), PSNR, as well as BD-rate savings~\cite{bjontegaard2001calculation}.

\paragraph{Runtime}
We measure the runtime of our transformers on multiple accelerators:
NVidia P100, V100, A100, 3090Ti GPUs, and Google Cloud TPUv3 and TPUv4.
We measure the $S$ forward passes required through the models, ignoring device transfer times, range coding, and the decoder $D$, since these are constant across all models. 
We report GFLOPS/image and milliseconds per image, where ``image'' means $2000{\times}1500$px.
For each accelerator, we chose the largest batch size that saturates it. For TPUs, we parallelize the model over 4 chips. 

\section{Results}

\begin{figure}[t]
\centering
\includegraphics[width=0.6\linewidth]{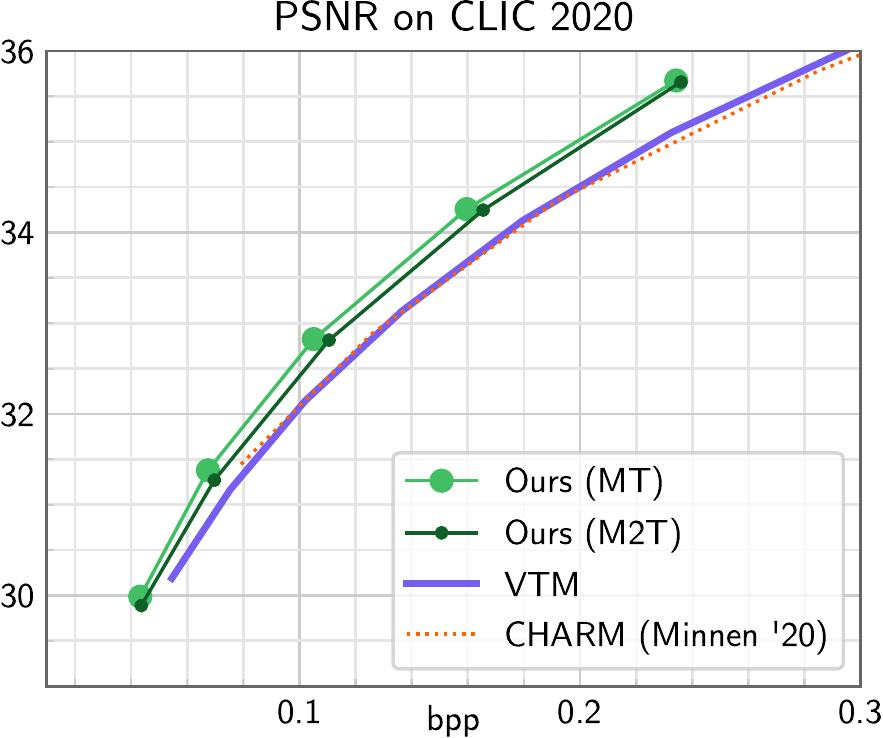}
\caption{\label{fig:rdclic}Rate distortion results on CLIC2020. }
\end{figure}

\paragraph{Rate-distortion}
In Fig.~\ref{fig:rd}, we compare rate-distortion performance on Kodak.
We can see that our model outperforms the previous state-of-the-art.
In Fig.~\ref{fig:rdclic}, we present results on CLIC2020 to show that also there, we significantly outperform the non-neural state-of-the-art VTM.
We use $S{=}12$ for \vanilla.
Sec.~\ref{app:sec:rawdata} provides the data underlying these plots.

\paragraph{Runtime}
In Fig.~\ref{fig:speed}, we compare inference speeds of our models on TPU v3/4 and 3090Ti/A100 GPUs.
In Sec~\ref{app:sec:speedall}, we also show P100, V100 and FLOPS.
Depending on the accelerator, \perm achieves 
$2.7{\times}-5.2{\times}$ practical wall clock speedups over the \vanilla model.
We note that \perm operates in the subsecond-per-large-image regime ($2000{\times}1500$px), putting it in the realm of practical image compression schemes.
\vanilla also achieves subsecond inference on a consumer GPU (3090Ti) if we use $S{=}8$.

However, we would also like to note that the runtime optimized channel-autoregressive and checkerboard-based convolutional entropy model in ELIC reports $46.06$ms for a $1890{\times}2048$px image on a consumer NVidia Titan XP~\cite[Table 3]{he2022elic}.
On the other end of the spectrum, the transformer-based ContextFormer reports full decoding speeds in the order of multiple minutes~\cite[Table 1]{koyuncu2022contextformer}.
As such, our contribution lies in developing a fast and simple transformer model that is largely based on the vanilla transformer encoder architecture used in BERT \cite{devlin2018bert} and ViT \cite{dosovitskiy2020image}, and still achieves state-of-the-art rate-distortion performance.

\paragraph{Certainty} We visualize the certainty of the entropy model of \vanilla in Fig.~\ref{fig:sampling} by sampling from the model multiple times and showing the sample mean. The underlaying samples are shown in Sec.~\ref{app:sec:samples}.

\subsection{Ablations} \label{sec:res:ablations}

\paragraph{Masking Schedules}

For the \vanilla model trained for $\lambda=0.00125$, we show the impact of various masking schedules in Fig.~\ref{fig:schedules}.
We see that using $\alpha>1$ is crucial to get low rates, but the gains start to saturate at around $\alpha=2.2$.
We see that for lower $\alpha$, the entropy-based masking schedule is optimal (top part of the lower right plot, lines with circle marker).
As we go towards the optimal $\alpha$, QLDS becomes the optimal schedule.
Finally, we see that increasing the number of autoregressive steps beyond 8 leads to limited gains if $\alpha>1$.

\paragraph{Architecture}

Our approach relies on standard components, so we only ablate the compression-specific choices: We explore using $C{=}320$ channels, since this is a common choice in the literature (\eg,~\cite{he2022elic,minnen2020channel}).
We also explore training with a single mixture instead of 3. The results are shown in Table~\ref{tab:abl}.

\begin{table}[t]
    \centering
    \begin{tabular}{lccc}
    \toprule
    Model & $C$ &  $N_M$ &  BD-rate savings\\
    \midrule
    \vanilla (default) &$192$ & 3 & -11.6\% \\
    \vanilla (more channels) &$320$ & 3 & -8.27\% \\
    \vanilla (one mixture) &$192$ & 1 & -7.95\% \\
    \bottomrule
    \end{tabular}
    \caption{\label{tab:abl} Ablating number of channels in the representation $C$ and number of mixtures $N_M$. Rate savings are over VTM (lower is better).}
\end{table}

\vfill
\section{Conclusion}

In this work, we made two significant contributions: We showed how a vanilla MaskGIT-like transformer can be applied to neural image compression to obtain state-of-the-art results at practical inference times on various accelerator platforms, without relying on a multi-scale model.
We also demonstrated that this model performs well with a fixed, deterministic schedule independent of the input, which allowed us to develop a second model class with masked attention, \perm. This model bridges between MaskGIT-like transformers and autoregressive transformers.

\paragraph{Acknowledgements}
We thank Pieter-Jan Kindermans and David Minnen for the insightful discussions.

\clearpage

{\small
\bibliographystyle{ieee_fullname}
\bibliography{egbib}

\begin{thebibliography}{10}\itemsep=-1pt

\bibitem{agustsson2022multi}
Eirikur Agustsson, David Minnen, George Toderici, and Fabian Mentzer.
\newblock Multi-realism image compression with a conditional generator.
\newblock {\em arXiv preprint arXiv:2212.13824}, 2022.

\bibitem{agustsson2019extreme}
Eirikur Agustsson, Michael Tschannen, Fabian Mentzer, Radu Timofte, and Luc~Van
  Gool.
\newblock Generative adversarial networks for extreme learned image
  compression.
\newblock In {\em The IEEE International Conference on Computer Vision (ICCV)},
  October 2019.

\bibitem{balle2020nonlinear}
Johannes Ball{\'e}, Philip~A Chou, David Minnen, Saurabh Singh, Nick Johnston,
  Eirikur Agustsson, Sung~Jin Hwang, and George Toderici.
\newblock Nonlinear transform coding.
\newblock {\em IEEE Journal of Selected Topics in Signal Processing},
  15(2):339--353, 2020.

\bibitem{balle2016code}
Johannes Ball{\'e}, Valero Laparra, and Eero~P Simoncelli.
\newblock End-to-end optimization of nonlinear transform codes for perceptual
  quality.
\newblock {\em arXiv preprint arXiv:1607.05006}, 2016.

\bibitem{balle2016end}
Johannes Ball{\'e}, Valero Laparra, and Eero~P Simoncelli.
\newblock End-to-end optimized image compression.
\newblock {\em arXiv preprint arXiv:1611.01704}, 2016.

\bibitem{balle2018variational}
Johannes Ball{\'e}, David Minnen, Saurabh Singh, Sung~Jin Hwang, and Nick
  Johnston.
\newblock Variational image compression with a scale hyperprior.
\newblock In {\em International Conference on Learning Representations (ICLR)},
  2018.

\bibitem{tfc_github}
Johannes Ballé, Sung~Jin Hwang, and Eirikur Agustsson.
\newblock {T}ensor{F}low {C}ompression: Learned data compression, 2022.

\bibitem{bjontegaard2001calculation}
Gisle Bjontegaard.
\newblock Calculation of average psnr differences between rd-curves.
\newblock {\em ITU SG16 Doc. VCEG-M33}, 2001.

\bibitem{blau2019rethinking}
Yochai Blau and Tomer Michaeli.
\newblock Rethinking lossy compression: The rate-distortion-perception
  tradeoff.
\newblock {\em arXiv preprint arXiv:1901.07821}, 2019.

\bibitem{chang2023muse}
Huiwen Chang, Han Zhang, Jarred Barber, AJ Maschinot, Jose Lezama, Lu Jiang,
  Ming-Hsuan Yang, Kevin Murphy, William~T Freeman, Michael Rubinstein, et~al.
\newblock Muse: Text-to-image generation via masked generative transformers.
\newblock {\em arXiv preprint arXiv:2301.00704}, 2023.

\bibitem{chang2022maskgit}
Huiwen Chang, Han Zhang, Lu Jiang, Ce Liu, and William~T Freeman.
\newblock Maskgit: Masked generative image transformer.
\newblock In {\em Proceedings of the IEEE/CVF Conference on Computer Vision and
  Pattern Recognition}, pages 11315--11325, 2022.

\bibitem{cheng2020learned}
Zhengxue Cheng, Heming Sun, Masaru Takeuchi, and Jiro Katto.
\newblock Learned image compression with discretized gaussian mixture
  likelihoods and attention modules.
\newblock In {\em Proceedings of the IEEE/CVF Conference on Computer Vision and
  Pattern Recognition}, pages 7939--7948, 2020.

\bibitem{devlin2018bert}
Jacob Devlin, Ming-Wei Chang, Kenton Lee, and Kristina Toutanova.
\newblock Bert: Pre-training of deep bidirectional transformers for language
  understanding.
\newblock {\em arXiv preprint arXiv:1810.04805}, 2018.

\bibitem{dosovitskiy2020image}
Alexey Dosovitskiy, Lucas Beyer, Alexander Kolesnikov, Dirk Weissenborn,
  Xiaohua Zhai, Thomas Unterthiner, Mostafa Dehghani, Matthias Minderer, Georg
  Heigold, Sylvain Gelly, et~al.
\newblock An image is worth 16x16 words: Transformers for image recognition at
  scale.
\newblock {\em arXiv preprint arXiv:2010.11929}, 2020.

\bibitem{el2022image}
Alaaeldin El-Nouby, Matthew~J Muckley, Karen Ullrich, Ivan Laptev, Jakob
  Verbeek, and Herv{\'e} J{\'e}gou.
\newblock Image compression with product quantized masked image modeling.
\newblock {\em arXiv preprint arXiv:2212.07372}, 2022.

\bibitem{esser2021taming}
Patrick Esser, Robin Rombach, and Bjorn Ommer.
\newblock Taming transformers for high-resolution image synthesis.
\newblock In {\em Proceedings of the IEEE/CVF conference on computer vision and
  pattern recognition}, pages 12873--12883, 2021.

\bibitem{vtm17}
Fraunhofer Gesellschaft.
\newblock {VTM 17.1}.
\newblock
  \url{https://vcgit.hhi.fraunhofer.de/jvet/VVCSoftware_VTM/-/releases/VTM-17.1},
  2022.

\bibitem{he2022elic}
Dailan He, Ziming Yang, Weikun Peng, Rui Ma, Hongwei Qin, and Yan Wang.
\newblock Elic: Efficient learned image compression with unevenly grouped
  space-channel contextual adaptive coding.
\newblock In {\em Proceedings of the IEEE/CVF Conference on Computer Vision and
  Pattern Recognition}, pages 5718--5727, 2022.

\bibitem{he2021checkerboard}
Dailan He, Yaoyan Zheng, Baocheng Sun, Yan Wang, and Hongwei Qin.
\newblock Checkerboard context model for efficient learned image compression.
\newblock In {\em Proceedings of the IEEE/CVF Conference on Computer Vision and
  Pattern Recognition}, pages 14771--14780, 2021.

\bibitem{flax2020github}
Jonathan Heek, Anselm Levskaya, Avital Oliver, Marvin Ritter, Bertrand
  Rondepierre, Andreas Steiner, and Marc van {Z}ee.
\newblock {F}lax: A neural network library and ecosystem for {JAX}, 2020.

\bibitem{kodakurl}
{Kodak PhotoCD dataset}.
\newblock \url{http://r0k.us/graphics/kodak/}, 2022.

\bibitem{koyuncu2022contextformer}
A~Burakhan Koyuncu, Han Gao, and Eckehard Steinbach.
\newblock Contextformer: A transformer with spatio-channel attention for
  context modeling in learned image compression.
\newblock {\em arXiv preprint arXiv:2203.02452}, 2022.

\bibitem{kuipers2012uniform}
Lauwerens Kuipers and Harald Niederreiter.
\newblock {\em Uniform distribution of sequences}.
\newblock Courier Corporation, 2012.

\bibitem{mentzer2018conditional}
Fabian Mentzer, Eirikur Agustsson, Michael Tschannen, Radu Timofte, and Luc
  Van~Gool.
\newblock Conditional probability models for deep image compression.
\newblock In {\em IEEE Conference on Computer Vision and Pattern Recognition
  (CVPR)}, 2018.

\bibitem{mentzer2022vct}
Fabian Mentzer, George Toderici, David Minnen, Sung-Jin Hwang, Sergi Caelles,
  Mario Lucic, and Eirikur Agustsson.
\newblock {VCT}: A video compression transformer.
\newblock {\em arXiv preprint arXiv:2206.07307}, 2022.

\bibitem{mentzer2020high}
Fabian Mentzer, George~D Toderici, Michael Tschannen, and Eirikur Agustsson.
\newblock High-fidelity generative image compression.
\newblock {\em Advances in Neural Information Processing Systems}, 33, 2020.

\bibitem{minnen2018joint}
David Minnen, Johannes Ball{\'e}, and George~D Toderici.
\newblock Joint autoregressive and hierarchical priors for learned image
  compression.
\newblock In {\em Advances in Neural Information Processing Systems}, pages
  10771--10780, 2018.

\bibitem{minnen2020channel}
David Minnen and Saurabh Singh.
\newblock Channel-wise autoregressive entropy models for learned image
  compression.
\newblock {\em arXiv preprint arXiv:2007.08739}, 2020.

\bibitem{qian2022entroformer}
Yichen Qian, Ming Lin, Xiuyu Sun, Zhiyu Tan, and Rong Jin.
\newblock Entroformer: A transformer-based entropy model for learned image
  compression.
\newblock {\em arXiv preprint arXiv:2202.05492}, 2022.

\bibitem{roberts2018unreasonable}
Martin Roberts.
\newblock {Unreasonable Effectiveness of Quasirandom Sequences}.
\newblock
  http://extremelearning.com.au/unreasonable-effectiveness-of-quasirandom-sequences/.
\newblock Accessed: 2023-01-13.

\bibitem{salimans2017pixelcnnpp}
Tim Salimans, Andrej Karpathy, Xi Chen, and Diederik~P Kingma.
\newblock Pixelcnn++: Improving the pixelcnn with discretized logistic mixture
  likelihood and other modifications.
\newblock {\em arXiv preprint arXiv:1701.05517}, 2017.

\bibitem{theis2017lossy}
Lucas Theis, Wenzhe Shi, Andrew Cunningham, and Ferenc Huszar.
\newblock Lossy image compression with compressive autoencoders.
\newblock In {\em International Conference on Learning Representations (ICLR)},
  2017.

\bibitem{tschannen2018deep}
Michael Tschannen, Eirikur Agustsson, and Mario Lucic.
\newblock Deep generative models for distribution-preserving lossy compression.
\newblock In {\em Advances in Neural Information Processing Systems}, pages
  5929--5940, 2018.

\bibitem{vaswani2017attention}
Ashish Vaswani, Noam Shazeer, Niki Parmar, Jakob Uszkoreit, Llion Jones,
  Aidan~N Gomez, {\L}ukasz Kaiser, and Illia Polosukhin.
\newblock Attention is all you need.
\newblock {\em Advances in neural information processing systems}, 30, 2017.

\bibitem{villegas2022phenaki}
Ruben Villegas, Mohammad Babaeizadeh, Pieter-Jan Kindermans, Hernan Moraldo,
  Han Zhang, Mohammad~Taghi Saffar, Santiago Castro, Julius Kunze, and Dumitru
  Erhan.
\newblock Phenaki: Variable length video generation from open domain textual
  description.
\newblock {\em arXiv preprint arXiv:2210.02399}, 2022.

\bibitem{williams1989learning}
Ronald~J Williams and David Zipser.
\newblock A learning algorithm for continually running fully recurrent neural
  networks.
\newblock {\em Neural computation}, 1(2):270--280, 1989.

\bibitem{xiangmimt}
Jinxi Xiang, Kuan Tian, and Jun Zhang.
\newblock Mimt: Masked image modeling transformer for video compression.
\newblock In {\em International Conference on Learning Representations}, 2022.

\bibitem{xiong2020layer}
Ruibin Xiong, Yunchang Yang, Di He, Kai Zheng, Shuxin Zheng, Chen Xing,
  Huishuai Zhang, Yanyan Lan, Liwei Wang, and Tieyan Liu.
\newblock On layer normalization in the transformer architecture.
\newblock In {\em International Conference on Machine Learning}, pages
  10524--10533. PMLR, 2020.

\bibitem{Yang2022a}
Y. Yang, S. Mandt, and L. Theis.
\newblock An introduction to neural data compression.
\newblock preprint, 2022.

\bibitem{zhu2021transformer}
Yinhao Zhu, Yang Yang, and Taco Cohen.
\newblock Transformer-based transform coding.
\newblock In {\em International Conference on Learning Representations}, 2021.

\bibitem{zou2022devil}
Renjie Zou, Chunfeng Song, and Zhaoxiang Zhang.
\newblock The devil is in the details: Window-based attention for image
  compression.
\newblock In {\em Proceedings of the IEEE/CVF Conference on Computer Vision and
  Pattern Recognition}, pages 17492--17501, 2022.

\end{thebibliography}
}

\clearpage
\appendix

\section{Appendix}

\subsection{Low discrepancy sequences}\label{app:sec:lds}

We reproduce the definition of discrepancy from Kuipers and Niederreiter~\cite[p.88]{kuipers2012uniform}. 
\begin{definition}
The \emph{discrepency} $D$ of a sequence of numbers $X=x_1, \dots, x_N$ is
\begin{equation}
D(X)=\sup_{0\leq\alpha<\beta\leq1}%
    \left|%
        \frac{A([\alpha,\beta);X)}{N}-(\beta-\alpha),
    \right|
\end{equation}
where $A(I;X)$ is the number of points in $X$ that fall into the interval $I$.
\end{definition}
Intuitively, this measures how ``dense'' the densest region in $X$ is compared to how many points we have overall. For example, a completely uniformly distributed $X$ with $N$ points has discrepancy $1/N$.

A ``low-discrepancy sequence'' (LDS) is a sequence where every subsequence has low discrepancy.
We use a LDS proposed by Roberts~\cite{roberts2018unreasonable}, with the following construction of the $i$-th point $\bm x_i$: 
\begin{align}
    \bm x_i = i \bm \phi \text{ mod } [1,1],
    \text{where } \bm \phi = [1/\rho, 1/\rho^2].
\end{align}
We refer to~\cite{roberts2018unreasonable} for details, but to give some intuition: in the above equation, the important property of $\bm \phi$ is that applying it repeatedly modulo 1 covers the space. Contrast, \eg, with $\bm \phi=[1,1]$, which would always map to the same point.

We note that this sequence is constructed for the unit cube, and we have to quantize it to use it for our representations. 
We do this by constructing the quantized LDS $\hat{\bm x_i}$ by scaling up and rounding the underlying ${\bm x_i}$ to the integer grid 
$\{0, \dots, \patchsize-1\}\times\{0,\dots,\patchsize-1\}$
and skipping over values of $i$ that are already used (since quantizing will lead to some values getting sampled more than once), \ie, we find the smallest integer $K$ such that quantizing 
 $\bm x_1, \dots, \bm x_K$ covers the $\patchsize{\times}\patchsize$ grid (which is $K=1381$ for $\patchsize=24$).

\subsection{Additional Accelerators}\label{app:sec:speedall}
\begin{figure}
\centering
\includegraphics[width=\linewidth]{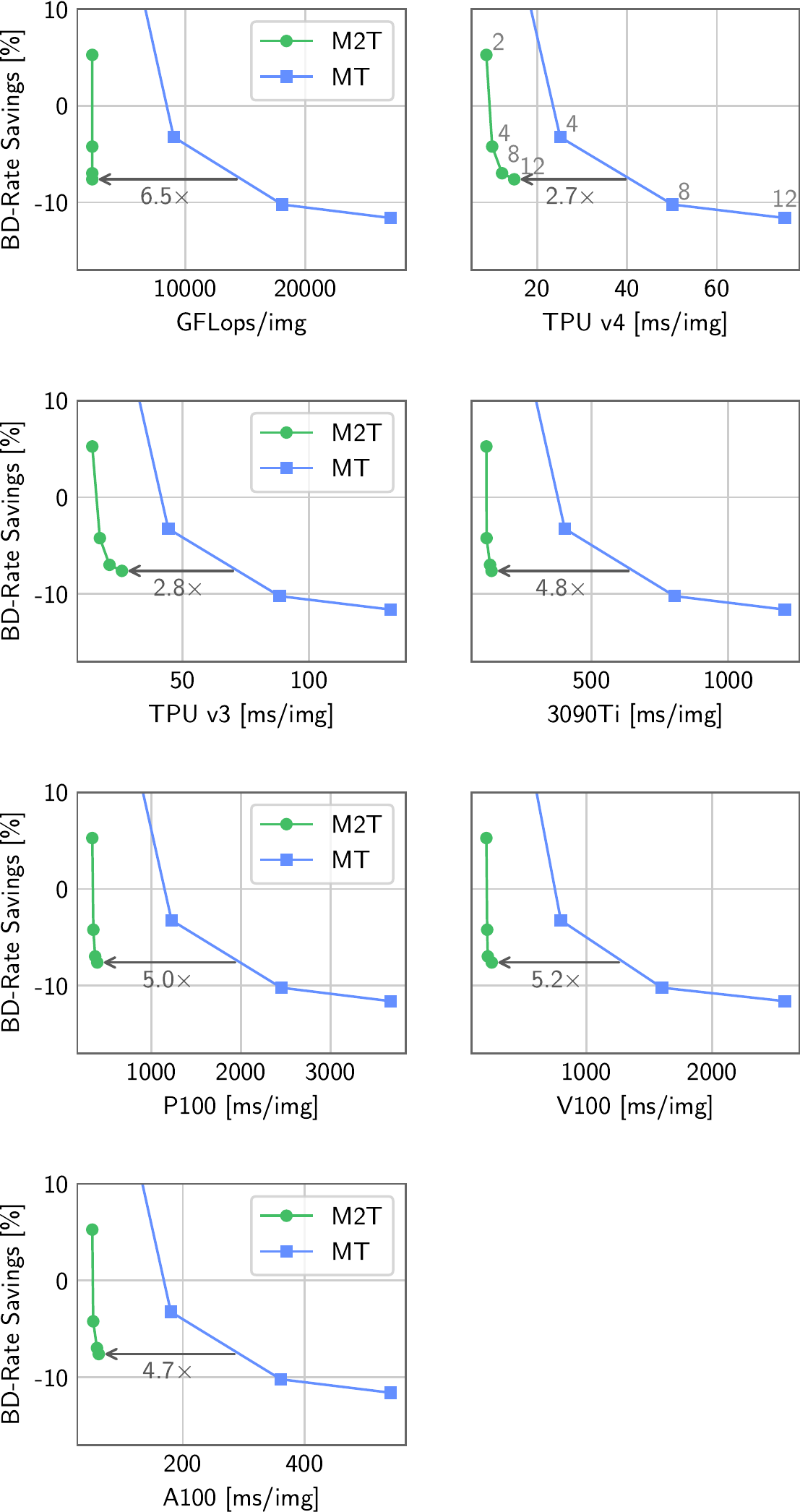}
\caption{\label{fig:speedall}Speed on all tested platforms as well as FLOPS.}
\end{figure}

We show runtime on all accelerators in Fig.~\ref{fig:speedall}.

\subsection{Implementation Details}\label{app:sec:details}

We use the ELIC architecture for the autoencoder as detailed in~\cite[Supplementary Material Sec 1.1, Table 1]{he2022elic}, where we use $M=256, N=192$.

We use the \texttt{AdamW} optimizer from Tensorflow Addons, using \texttt{weight\_decay}${=}0.03\cdot \text{learning rate}$,
$\beta_1{=}0.9,\beta_2{=}0.98, \epsilon{=}1\textsc{E}^{-9}$, \texttt{global\_clipnorm}${=}1$.\footnote{\url{https://www.tensorflow.org/addons/api_docs/python/tfa/optimizers/AdamW}}

To improve the training stability of the GMM, we found it important to adopt the Laplace tail mass approach~\cite{tfc_github} (with weight $0.001$), falling back to to the (more numerically stable) Laplacian distribution when the probability predicted by GMM vanishes.

Previous works (\eg~\cite{minnen2018joint, minnen2020channel}) center the representation with the mean prediction of the entropy model during quantization, 
which couples entropy modelling and the reconstruction.
For our models, this poses a problem, since during training, we do not know the mean before doing one pass through the model, so  we would have to double the forward passes to do this.
To simplify the setup, we simply drop this.
As an important benefit, dropping the centering also allows us to employ a patched inference scheme described next without creating block artifacts in the reconstruction.

\subsection{Algorithms}

We show pseudo-code for how our masked transformers can be used on the sender side in neural compression in Als~\ref{alg:send:vanilla}~\ref{alg:send:perm}.

\begin{figure}[b]

\begin{algorithm}[H]
\caption{\label{alg:send:vanilla}\vanilla Sender}
\centering
\begin{algorithmic}
\Require Input $y$ of shape $(b', \patchsize^2, c)$, seed $s$.
\State $c \gets ones(S, C) \cdot \text{mask token}$ \Comment{Current Input}
\State masker $\gets$ make\_masker(seed=$s$)
\For{$i \in \{1,\dots,\text{masker.num\_steps}\}$}
    \State params $\gets$ transformer($c$)
    \State $m$ $\gets$ masker.get\_mask(params)
    \State bit\_stream $\gets$ bitencode($y[:, m, :]$, params$[:, m, :]$)
    \State $c \gets y[:, m, :]$  \Comment{Uncover input}
\EndFor
\end{algorithmic}
\end{algorithm}
\begin{algorithm}[H]
\caption{\label{alg:send:perm}\perm Sender}
\centering
\begin{algorithmic}
\Require Input $y$ of shape $(b', P^2, C)$, seed $s$.
\State masker $\gets$ make\_masker(seed=$s$)
\State $y_\text{perm}\gets$ permute(masker, $y$)
\State \textcolor{code}{t $\gets$ 0}  \Comment{Total Uncovered}
\State $c \gets []$ \Comment{Current Input}
\For{$i \in \{1,\dots,\text{masker.num\_steps}\}$}
\color{code}
    \State $m_\text{len}$ $\gets$ len(masker.get\_mask(params))
    \State $c$.extend(ones(b, $m_\text{len}$, C) $\cdot$ mask\_token
\color{black}
    \State params, \textcolor{code}{cache} $\gets$ transformer($c$, \textcolor{code}{cache})
\color{code}
\color{black}
    \State bit\_stream $\gets$ bitencode(\\
      \hspace{3em}\textcolor{code}{%
      $y_\text{perm}[:, t:t+m_\text{len}, :]$}, params)
    \State $c$.extend(\textcolor{code}{$y_\text{perm}[:,t:t+m_\text{len},:]$})  \Comment{Uncover input}
\EndFor
\end{algorithmic}
\end{algorithm}

\end{figure}

\subsection{Samples}\label{app:sec:samples}
\begin{figure}
    \centering
    \includegraphics[width=\linewidth]{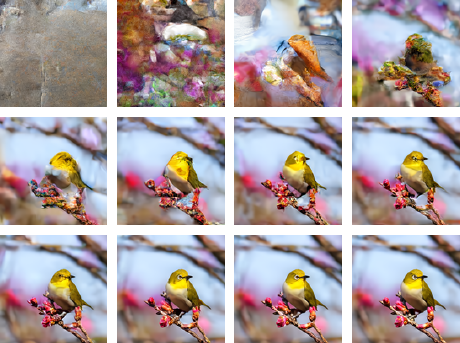}
    \caption{\label{fig:rawsamples}Raw samples.}
\end{figure}

We show samples from the entropy model in Fig.~\ref{fig:rawsamples}.
For Fig.~\ref{fig:sampling} in the main text, we take the average over 50 samples per step.

\subsection{Flax Implementation}\label{app:sec:flax}

We show the flax implementation of \perm in Fig.~\ref{lst:attn}.

\definecolor{codegreen}{rgb}{0,0.6,0}
\definecolor{codegray}{rgb}{0.5,0.5,0.5}
\definecolor{codepurple}{rgb}{0.58,0,0.82}

\lstdefinestyle{mystyle}{
    commentstyle=\color{codegreen},
    keywordstyle=\color{magenta},
    numberstyle=\tiny\color{codegray},
    stringstyle=\color{codepurple},
    basicstyle=\ttfamily\scriptsize,
    breakatwhitespace=false,         
    breaklines=true,                 
    captionpos=b,                    
    keepspaces=true,                 
    showspaces=false,                
    showstringspaces=false,
    showtabs=false,                  
    tabsize=2
}

\lstset{style=mystyle}

\begin{figure*}
\centering
\begin{minipage}[c]{0.9\linewidth}%
\small
\begin{lstlisting}[language=Python]
import functools
from typing import Optional
from flax.linen import attention
from flax.linen.linear import DenseGeneral
from flax.linen.module import compact, merge_param
from jax import lax
import jax.numpy as jnp

# Forked from flax.
class PermutedMHA(attention.MultiHeadDotProductAttention):
  @compact
  def __call__(
      self, inputs_q, inputs_kv, mask=None, deterministic: Optional[bool] = None):
    if self.dropout_rate > 0.0:
      raise NotImplementedError('Dropout removed for simplicity.')
    if len(mask.shape) != 2:
      raise NotImplementedError('Only supporting 2D masks at the moment!')
    features = self.out_features or inputs_q.shape[-1]
    qkv_features = self.qkv_features or inputs_q.shape[-1]
    assert qkv_features % self.num_heads == 0
    head_dim = qkv_features // self.num_heads
    dense = functools.partial(
        DenseGeneral, axis=-1, dtype=self.dtype, param_dtype=self.param_dtype,
        features=(self.num_heads, head_dim), kernel_init=self.kernel_init, bias_init=self.bias_init,
        use_bias=self.use_bias, precision=self.precision)
    query, key, value = (
        dense(name='query')(inputs_q), dense(name='key')(inputs_kv), dense(name='value')(inputs_kv))
    if self.decode:
      assert mask is not None
      is_initialized = self.has_variable('cache', 'cached_key')
      cached_key = self.variable('cache', 'cached_key', jnp.zeros, key.shape, key.dtype)
      cached_value = self.variable('cache', 'cached_value', jnp.zeros, value.shape, value.dtype)
      cache_index = self.variable('cache', 'cache_index', lambda: jnp.array(0, dtype=jnp.int32))
      if is_initialized:
        *batch_dims, max_length, num_heads, depth_per_head = cached_key.value.shape
        *_, count, query_num_heads, query_depth_per_head = query.shape
        if query_num_heads != num_heads or query_depth_per_head != depth_per_head:
          raise ValueError("Invalid dimensions")
        cur_index = cache_index.value
        start_indices = (0,) * len(batch_dims) + (cur_index, 0, 0)
        key = lax.dynamic_update_slice(cached_key.value, key, start_indices)
        value = lax.dynamic_update_slice(cached_value.value, value, start_indices)
        cached_key.value = key
        cached_value.value = value
        # We just slice the mask, since it's q_length, kv_length, and we slide through Q.
        mask = lax.dynamic_slice(mask, (cur_index, 0), (count, max_length))
        cache_index.value = cache_index.value + count
    x = self.attention_fn(
        query, key, value,
        mask=mask, dropout_rng=None, dropout_rate=self.dropout_rate,
        broadcast_dropout=self.broadcast_dropout,
        deterministic=True, dtype=self.dtype, precision=self.precision)  
    return DenseGeneral(
        features=features, axis=(-2, -1), kernel_init=self.kernel_init,
        bias_init=self.bias_init, use_bias=self.use_bias, dtype=self.dtype,
        param_dtype=self.param_dtype, precision=self.precision, name='out')(x)

\end{lstlisting}
\end{minipage}
\caption{\label{lst:attn}Flax attention implementation for \perm.}
\end{figure*}

\subsection{Raw Data}\label{app:sec:rawdata}

We provide the raw data for the figures of the main text in Table~\ref{tab:data:kodak},~\ref{tab:data:clic}.

\begin{table}[]
    \centering
\begin{tabular}{cccc}
\toprule
  \multicolumn{2}{c}{\vanilla} &
  \multicolumn{2}{c}{\perm} \\
        0 &          1 &         2 &          3 \\
\midrule
 0.058108 &  27.079653 &  0.059169 &  27.034135 \\
 0.094242 &  28.468065 &  0.097261 &  28.408479 \\
 0.153969 &  29.985175 &  0.162184 &  29.978115 \\
 0.247314 &  31.652337 &  0.257729 &  31.644149 \\
 0.380635 &  33.393239 &  0.385378 &  33.372010 \\
\bottomrule
\end{tabular} 
\caption{\label{tab:data:kodak}Raw data for Fig.~\ref{fig:rd} (Rate-distorion on Kodak).}
\end{table}

\begin{table}[]
    \centering
\begin{tabular}{cccc}
\toprule
  \multicolumn{2}{c}{\vanilla} &
  \multicolumn{2}{c}{\perm} \\
        0 &          1 &         2 &          3 \\
\midrule
 0.043276 &  29.986596 &  0.043651 &  29.888452 \\
 0.067407 &  31.379083 &  0.069709 &  31.270813 \\
 0.105127 &  32.825032 &  0.110558 &  32.813387 \\
 0.159596 &  34.256665 &  0.165478 &  34.246637 \\
 0.234286 &  35.674083 &  0.235964 &  35.657870 \\
\bottomrule
\end{tabular}
\caption{\label{tab:data:clic}Raw data for Fig.~\ref{fig:rdclic} (Rate-distortion on CLIC)}
\end{table}

\end{document}